\documentclass[11pt,reqno]{article}
\evensidemargin\oddsidemargin

\usepackage{latexsym}
\usepackage{graphicx}
\usepackage{amssymb}
\usepackage{amsmath}
\usepackage{amsthm}
\usepackage{url}
\usepackage[scaled=.92]{helvet}  
\newcommand{\ctr}[1]{\begin{center} #1 \end{center}}
\newcommand{\oo}{\infty}

\newcommand{\alx}[1]{\begin{align*} #1 \end{align*}}
\newcommand{\RR}{\mathbf{R}}

\newcommand{\NN}{\mathbf{N}}

\newcommand{\x}{\mathbf{x}}
\renewcommand{\L}{\mathfrak{L}}
\newcommand{\prf}{\noindent {\em Proof.}  }
\newcommand{\QED}{\hfill $\qed$}
\newcommand{\dom}{\mbox{\rm dom}}
\newcommand{\bin}{\mbox{bin}}

\newcommand{\floor}[1]{\lfloor #1 \rfloor}

\newcommand{\fdom}{\mbox{\rm \footnotesize{dom}}}
\newcommand{\fbin}{\mbox{\rm \footnotesize{bin}}}
\newcommand{\COM}[2]{{#1 \choose #2}}

\newtheorem{thm}{Theorem}
\newtheorem{cor}[thm]{Corollary}
\newtheorem{fact}[thm]{Fact}
\newtheorem{lem}[thm]{Lemma}
\newtheorem{example}[thm]{Example}
\newtheorem{defn}[thm]{Definition}
\newtheorem{prop}[thm]{Proposition}
\newtheorem{sch}[thm]{Scholium}

\begin{document}
\title{\bf Natural Halting Probabilities, Partial Randomness, and Zeta Functions}
\author{Cristian S. Calude$^{1}$
and Michael A. Stay$^{2}$\\
\phantom{xx}\\
$^{1}$Department of Computer Science\\ University of
Auckland, New Zealand\\ Email: {\tt
cristian@cs.auckland.ac.nz}\\
\phantom{xx}\\
$^{2}$Department of Mathematics\\
University of California Riverside, USA\\
Email: {\tt
mike@math.ucr.edu}}
\date{}
\maketitle

\thispagestyle{empty}
\begin{abstract}
We introduce  the  {\it zeta number, natural halting probability} and    {\it natural complexity} of a Turing machine and we relate them to Chaitin's Omega number,  halting probability, and program-size complexity. A classification of Turing machines according to their zeta  numbers is proposed: divergent, convergent and tuatara. We prove the existence of universal convergent and tuatara machines. Various results on (algorithmic) randomness and partial randomness are proved. For example, we show that the zeta number of a universal tuatara machine is c.e.\ and random.  A new type of partial randomness, asymptotic randomness,  is introduced. Finally we show that in contrast to classical (algorithmic) randomness---which cannot be naturally characterised in terms of plain complexity---asymptotic randomness admits such a characterisation.
\end{abstract}

\maketitle

\section{Introduction}

We introduce  the  {\it zeta number, natural halting probability} and    {\it natural complexity} of a Turing machine and we relate them to Chaitin's Omega number,  halting probability, and program-size complexity.   A classification of Turing machines according to their zeta numbers is proposed: divergent (zeta number is infinite), convergent (zeta number is finite), and tuatara (zeta number is less than  or equal to one). Every self-delimiting Turing machine is tuatara, but the converse is not true. Also, there exist universal convergent and tuatara machines; there is a tuatara machine universal for the class of convergent machines.

The zeta number of a universal self-delimiting Turing machines is c.e.\ and (algorithmically) random, and for each  tuatara machine  there effectively exists a  self-delimiting Turing machine whose Chaitin halting probability equals its zeta number; if the tuatara machine is universal, then the self-delimiting Turing machine can also be taken to be universal.

For each self-delimiting Turing machine there is a  tuatara machine whose zeta number is exactly the Chaitin halting probability of the self-delimiting Turing machine; it is an open problem whether the tuatara machine can be chosen to be a universal self-delimiting Turing machine in the case when the original machine is universal.

Let $s>1$ be a computable real, $T$ a universal Turing machine, and $K_{T}$ be the plain complexity induced by $T$. 
In analogy with the notion of Chaitin partially random reals we introduce  the notion of a  ``$1/s$-$K$-random  real'' (a real $\alpha = 0.x_{1}\cdots x_{m}\cdots$ such that the prefixes of its binary expansion  are  $1/s-K$-random, i.e.\  $K_{T}(x_{1}\cdots x_{m}) \ge m/s-c$, for some $c\ge 0$ and all $m\ge 1$) as well as the notion of  an ``asymptotically random real'' ($1/t$-$K$-random real, for every computable $t>s>1$).  

The result due to Chaitin and Martin-L\" of showing that 
there is no infinite sequence whose prefixes  have all  {\it maximal}  $K$ complexity  (also true for $H$ complexity)
 is no longer  true  for asymptotically random reals (Theorem~\ref{h=k} and Theorem~\ref{nongregml}). However,   $1/s-K$-randomness is different from Chaitin $1/s-$randomness (Proposition~\ref{hnot=k}).  Every c.e.\ random number is asymptotically random (Theorem~\ref{omegaAKrand}), but the converse implication fails to be true: there exists a self-delimiting Turing machine whose zeta number is asymptotically random, but not random 
(Theorem~\ref{arnotrand}).

\if01
The result due to Chaitin and Martin-L\" of showing that the plain complexity $K$ cannot characterise random reals is no longer  true  for asymptotically random reals (Theorem~\ref{h=k}), but  $1/s-K$-randomness is different from Chaitin $1/s-$randomness (Proposition~\ref{hnot=k}).  Every c.e.\ random number is asymptotically random (Theorem~\ref{omegaAKrand}), but the converse implication fails to be true: there exists a self-delimiting Turing machine whose zeta number is asymptotically random, but not random 
(Theorem~\ref{arnotrand}).
\fi
Various examples illustrate the above notions and results.
Some open problems conclude the paper.

\section{Omega and zeta numbers}

It is well-known that the Halting Problem, i.e.\ the problem of deciding whether an arbitrary Turing machine halts or not on a given input, is Turing uncomputable. The probabilistic version of the Halting Problem, first studied by Chaitin \cite{chaitin75,chaitin87},  deals with the halting probability, i.e.\  the probability that an arbitrary Turing machine halts on a  randomly chosen input. Chaitin's  halting probability was studied intensively by various authors (see \cite{solovaymanu,RozenbergSalomaa,cris02,dh}).  Chaitin's halting probability is not defined for every Turing machine, hence Chaitin and his followers have worked with a sub-class of Turing machines which has equal enumeration power as the class of all Turing machines, namely the self-delimiting Turing machines.

A {\em self-delimiting  Turing machine} $C$ is a Turing machine  which processes binary strings into binary strings and has a {\it prefix-free}  domain; that is, if $C(x)$ halts (is defined) and $y$ is either a proper prefix or a proper extension of $x$, then $C(y)$ is not defined. The domain of $C$, $\dom (C)$, is the set of strings on which $C$ halts (is defined).

\medskip

\begin{defn}[Chaitin's Omega Number] {\rm The} halting probability (Omega Number) {\rm of a self-delimiting Turing machine $C$ is }
\[ \Omega_{\fdom (C)} = \sum_{p\;\in\;\fdom(C)}2^{-|p|}. \]
\end{defn}

The number $\Omega_{\fdom (C)}$, usually written  $\Omega_{C}$, is a halting probability. Indeed, pick, at random using the Lebesgue measure on $[0,1]$, a real $\alpha$ in the unit interval and note that the probability that some initial prefix of the binary expansion of $\alpha$ lies in the prefix-free set $\dom(C)$ is exactly  $\Omega_C$.
\medskip

More formally, let $\Sigma =\{0,1\}$ and let  $\Sigma^*, \Sigma^{\omega}$ be the set of binary strings and infinite binary sequences, respectively. For 
$A \subseteq \Sigma^*, A\Sigma^{\omega}=\{w\x\ \mid \ w\in A, \ \x\in\Sigma^\omega\}$, the cylinder induced by $A$, 
is the set of sequences having a prefix in $A$. The sets $A\Sigma^\omega$ are the open sets  in the natural topology on $\Sigma^\omega$. Let $\mu$ denote the usual product measure on $\Sigma^\omega$  given by the uniform distribution $\mu(\{0\}\Sigma^\omega)=\mu(\{1\}\Sigma^\omega) = 2^{-1}$.  For a measurable set {\bf C} of infinite sequences, $\mu({\bf C})$ is the probability that $\x\in {\bf C}$ when $\x$ is chosen by a random experiment in which an `independent toss of a fair coin' is used to decide whether $x_n=1$. If $A$ is prefix-free, then $\mu(A\Sigma^\omega)=\sum_{w\in A} 2^{-|w|} =  \Omega_A$; here $|w|$ is the length of the string $w$.  We assume everywhere that $\min\emptyset=\infty$. For more details see \cite{cris02,dh}.

Let   $\alpha = 0.x_{1}x_{2}\cdots x_{n}\cdots \in [0,1]$ with $x_{i}\in \{0,1\}$, and let $x_{1}x_{2}\cdots x_{n}\cdots$ be the unending binary expansion of $\alpha$. We put   $\alpha [n]= x_{1}x_{2}\cdots x_{n}$. If $y= y_{1}y_{2}\cdots y_{n}$, then $0.y = \sum_{i=1}^{n}y_{i}2^{-i}$.

\medskip

\begin{defn}{\rm  The Turing machine $U$ is {\it universal}  for a class $\Re$ of  Turing machines  if for
every Turing machine $C\in \Re$ there exists a fixed constant $c\ge 0$ (depending upon $U$ and $C$) such that for every $x\in \dom (C)$ there is a string $p_{x}\in \dom (U)$ with $|p_{x}|\le |x| +c$ and $U(p_{x})=C(x)$.  In case $U\in \Re$, we simply say that {\it the machine $U\in \Re$} is {\it universal}.}
\end{defn}

\medskip

A classical result states:

\smallskip

\begin{thm}
\label{univsdTM}{\rm \cite{chaitin75}}
 We can effectively construct a universal self-delimiting  Turing machine.
\end{thm}

\medskip

\begin{defn}
{\rm a)  The} plain complexity {\rm of the string $x\in\Sigma^{\ast}$ with respect to a  Turing machine $M$ is} $K_{M}(x)=\min\{|w|\mid w\in\Sigma^{\ast},\ M(w)=x\}.$

{\rm b) The} program-size complexity {\rm of the string $x\in\Sigma^{\ast}$ with respect to a self-delimiting Turing machine  $C$ is} $H_{C}(x)=\min\{|w|\mid w\in\Sigma^{\ast},\ C(w)=x\}.$
\end{defn}

\medskip

\begin{defn}{\rm a)  \cite{soare} A real $\alpha \in (0,1)$ is} computably enumerable {\rm(}c.e.{\rm ) if it is the  limit of  an increasing computable sequences of rationals. }

{\rm b) (\cite{tadaki,cst}) Let $\varepsilon$ be a computable real and $U$ a universal self-delimiting Turing machine. A real $\alpha \in (0,1)$ is Chaitin} $\varepsilon$-random {\rm if there is a constant $c$ such that for each $n\ge 1$, $H_{U} (\alpha [n]) \ge \varepsilon \cdot n-c$. We say that $\alpha$ is {\it Chaitin partially random} if it is} Chaitin $\varepsilon-$random {\rm for some computable real }  $1>\varepsilon>0$.

{\rm c) \cite{chaitin75}  A real $\alpha \in (0,1)$ is} (algorithmically) random {\rm if it is 1-random}, {\em i.e.\ there exists $c\ge 0$ such that for all } $m\ge 1$, $H_{U}(\alpha[m])\ge m-c$.
\end{defn}

\medskip

The following theorem gives a full characterisation of  c.e.\ and random reals:

\begin{thm}
\label{cerand=haltprob}{\rm (\cite{chkw,KS,cris02})}
 A real $\alpha \in  (0,1)$ is c.e.\ and random iff there exists a universal self-delimiting  Turing machine $U$ such that $\alpha = \Omega_{U}$.
\end{thm}

The definition of Chaitin's halting probability allows an apparent ``ambiguity'' as strings with the same length in the domain of the self-delimiting Turing machine contribute equally towards the  halting probability.\footnote{The ``ambiguity'' is apparent because from the first $n$ bits of $\Omega_{U}$ we effectively calculate the strings in $\dom (U)$ that determine these digits.} This motivates us to introduce a slightly different ``halting probability'' in which different strings in the domain of the machine have different contributions to the ``halting probability''. 

\medskip

Let $\NN = \{1,2,\ldots\}$ and let $\bin:\NN \rightarrow \Sigma^*$ be the bijection which associates to every $n\ge 1$ its  binary expansion  without the leading 1,    
\ctr{\begin{tabular}{r|c|c|c}
$n$&$n_2$&$\bin(n)$& $|\bin(n)|$\\
\hline
1&1&$\lambda$ & 0\\
2&10&0 & 1\\
3&11&1& 1\\
4&100&00 & 2\\
$\vdots$&$\vdots$&$\vdots$&$\vdots$
\end{tabular}}

\medskip

If $A\subset \Sigma^{*}$, then we define $ \Upsilon[A]=\{n\in \NN \mid \bin(n)\in A\}$. In other terms, the binary expansion of $n$ is $n_{2} = 1\bin(n)$.
\medskip

\begin{defn}[Zeta number of a Turing machine] {\rm The} zeta number  of the Turing machine $M$, {\rm denoted $\zeta_M$,  is }
\[\zeta_M = \sum_{n \in \Upsilon[\fdom(M)]} \frac{1}{n}\raisebox{.6ex}.\]
\end{defn}

\if01
The number  $\zeta_M$ may be thought as a ``halting probability''  for $M$ in the following sense: if there were such a thing as $e$-ary string, where $e$ is the base of the natural logarithm, then it would be the natural topology on $e$-ary sequences, just as $\Omega_M$ uses the binary topology.  Rather than strings of bits, we have strings of {\em nats}, where one nat conveys $1/\log(2)\simeq 1.44$ bits of information.  Sequences of nats lie somehere between binary and ternary sequences. 

Begin with ternary sequences, i.e., sequences over $\{0,1,2\}$. 
Intuitively, we have a coin with a wide edge, so some of the time
it lands on the edge as well.  The coin is fair if we count the edge and heads
together, but not heads only.  To make the coin binary, i.e. to force it to land only on heads or tails, one can trim off some of the edge after each time it lands ``heads".
\medskip

More formally, we make $n$ flips and consider those with no edge appearing.  The distribution is the following: begin with $k_0=1$.  At the $i$th flip $f_i$,
\alx{
P(f_i=0)&=k_i/2k_i = 1/2, \\
P(f_i=1)&=k_i/(2k_i+1),\\
P(f_i=2)&=1/(4k_i+2),
}
\[ k_{i+1}=
\begin{cases}
  2k_i,  \text{ if }f_i=0,\\
  2k_i+1, \text{ if }f_i=1.\\
\end{cases}
\]
As $n$ approaches infinity, we are essentially doing numerical integration:
\[ \lim_{n\rightarrow \infty}\sum_{i=2^n}^{2^{n+1}-1} 1/i = \int_1^2 1/x\; dx = \log 2.\]
If we were in base $e$, this integral, representing the probability that some event occurs, would be 1, and we would have a proper probability space.  
\medskip
\fi

The number $\zeta_M$ will be shown to be random in the same sense as
$\Omega_M$ in case $M$ is `universal' (for example, if $M$ is a universal self-delimiting Turing machine, Theorem~\ref{naturalhalt}). 

One might ask whether  there is also some sense in which
$\zeta_M$ is a halting probability. 
For many Turing machines, $\zeta_M$ is not a probability; for example, a total Turing machine $M$, i.e. $\dom (M) = \Sigma^{*}$, has $\zeta_M=\infty$.

However, for a universal self-delimiting Turing machine $M$,  $\zeta_M$ is a halting probability. Here is an informal argument. In an alphabet with $k$ symbols,
the probability that the $k$-ary expansion of $n$ appears is
proportional to
$ k^{-\floor{\log_k n}-1},$
while the measure assigned to $n$ in the definition of $\zeta_M$ is
$ k^{-\log_k n}.$
By letting $k$ approach 1 from above, we can eliminate the roughness
in the measure due to the least integer function.  Fractional bases
$k$ correspond to strings in base $\lceil k \rceil$ with restrictions.  For
instance, using the golden ratio $\phi = \frac{1+\sqrt{5}}{2} \approx
1.618$ as a base, we get the ``Fibonaccimal'' \cite{Fib} expansion.
Here, numbers are represented by binary strings in which consecutive 1 digits
are prohibited.  As $k$ approaches 1, the measure of $n$ approaches
$1/n$.

\medskip

\begin{defn} {\rm (Zeta classification of Turing machines). According to the {\it zeta number}, Turing machines can be classified into the following three classes:
\begin{itemize}
\item {\it zeta divergent Turing machines}: those machines $M$ for which  $\zeta_M = \infty$,
\item {\it zeta convergent Turing machines}: those machines $M$ for which  $\zeta_M < \infty$,
\item {\it tuatara machines}\footnote{We
chose this name to commemorate the fact that the work was done in New
Zealand. Tuatara (``peaks on the back" in Maori) is a reptile (not a lizard)  found only in New Zealand. Tuatara is the last remaining member of the ancient group of reptiles {\it Sphenodontia}, the only survivor of a large group of reptiles that roamed the
earth at the time of dinosaurs. Tuatara has not changed its form much in over 225 million years!  Its relatives died out about 60 million years ago.  Tuatara has a `third eye'; its main role is to soak up ultraviolet rays in the first few months of life.  See more in \cite{tuatarawww}.}: those machines $M$ for which  $\zeta_M \le 1$.
\end{itemize}
}
\end{defn}

\medskip

\begin{prop}
\label{zetaprefixmachine}
Every self-delimiting Turing machine is a tuatara machine. More precisely, for every self-delimiting Turing machine $C$,
$\zeta_C$  is c.e.\ and $$1\ge\Omega_C \ge \zeta_C \ge \Omega_C/2\ge 0.$$
\end{prop}
\prf It is easy to see that $
\zeta_C$ is c.e.\ and
\alx{
1\ge  \Omega_C &\geq \sum_{n \in \Upsilon[\fdom(C)]  } 2^{-|\fbin(n)|} \\
           &= \sum_{n \in \Upsilon[\fdom(C)]  } 2^{-\floor{\log_2(n)}} \geq \sum_{n \in \Upsilon[\fdom(C)]  } 2^{-\log_2(n)}   = \zeta_C\\
           & \ge \sum_{n \in \Upsilon[\fdom(C)] } 2^{-\floor{\log_2(n)}-1}
          = \sum_{n \in \Upsilon[\fdom(C)] } 2^{-|\fbin(n)|-1} \\
          &   = \Omega_C/2 \ge 0.}
\\[-5ex]
\phantom{xxx}
\QED

\medskip

\if01
One particularly simple self-delimiting Turing machine is  Barker's language Iota \cite{BI}.  The simplest way to define Iota is in terms of Church's $\lambda-$calculus: the universal basis $\{S=\lambda xyz.xz(yz), K=\lambda xy.x\}$ suffices to produce every lambda term, but for universality it is not necessary to have two combinators.  There are one-combinator bases, known as {\em universal combinators}.   Iota is a very simple universal combinator, $\lambda f.fSK$, denoted $0$.  To make Iota unambiguous, there is a prefix operator, $1$, for application.

The construction is essentially a very stripped-down version of LISP with only one atom, 0; since the atom takes a single input, we can represent the open parenthesis with 1, and we note that closing parentheses are unnecessary.

Valid Iota programs are pre-order traversals of full binary trees.  The number of full binary trees with $n$ leaves is $C_{n-1}$, the $(n-1)$st Catalan number ($C_n \sim \frac{4^n}{\sqrt{\pi n^3}}$), and any traversal of such a tree with $n$ leaves will be $2n-1$ bits long.  Then, the sum over all trees using the natural measure is 1:

\begin{equation}
\label{iota=1}
\sum_{n>0} \frac{C_{n-1}}{2^{2n-1}} = 1. 
\end{equation}

That is, every infinite binary sequence (except for a set with measure zero) begins with the pre-order traversal of some full binary tree.

\medskip
\fi

We continue with the following 
result \cite{St06}:

\begin{thm}\label{density}
Let $U$ be a universal self-delimiting Turing machine. Then,
\[ \liminf_{n\rightarrow \infty} \, \frac{1}{n} \log (\#\{p \in \dom (U) \mid |p|\le n \}) = 1.\]
\end{thm}

\prf If $M$ is a one-to-one (as a partial function), self-delimiting Turing machine, then in view of the universality of $U$ we have: $\#\{q\in \dom(M) \mid |q| \le n-c\} \le \#\{p\in \dom(U) \mid |p| \le n\}$. To obtain the formula in the statement of the theorem we can choose $M$ such that $\dom(M)=\L$,  the Lukasiewicz language defined by the equation $\L = 0 \cup 1\cdot \L^{2}$ (see \cite{Kuich}); 
so,   for every odd $n$ we have
  $$ \#\{ q \in \dom(M)  \mid  |q| \le n\} = \sum_{{\rm odd   }  \phantom{x}  i}^n  \frac{1}{2i+1} \COM{2i+1}{i}
 = \sum_{{\rm odd   } \phantom{x}  i}^n C_i,$$ \noindent  where $C_{i}$ is the $i$th Catalan number (see \cite{Kuich}).

\QED

\medskip

\begin{fact}
\label{integerpower}
The domain of a universal self-delimiting Turing machine $U$ cannot be a set of strings such that every element has a length that is an integer power of two.
\end{fact}

\prf The result follows from Theorem~\ref{density}: otherwise, $\liminf_{n\rightarrow \infty} \, \frac{1}{n} \log (\#\{p \in \dom (U) \mid |p|\le n \}) \le 1/2$.

\QED

\if01
In view of the universality, $U$ simulates Iota with constant $c$.
Hence,   for all $p \in \dom(Iota)$, there exists a string $q_p$ such that $U(q_p) = Iota(p)$ and $|q_p|\leq |p|+c$.  Note that  all strings of the form $11010100x$ halt on Iota, where $x$ is a pre-order traversal of any full binary tree; this is equivalent to the lambda term $K(x)$.  If we assume that the domain of $U$ consists 
solely of strings whose preimage under $\bin$ is a
 power of two,
then, even ignoring the restriction that halting programs cannot be prefixes or extensions of each other, there are only
\[ \sum_{i=0}^{\floor{\log_2(c+8+|x|)}}2^i = 2^{\floor{\log_2(c+8+|x|)}+1} = \mbox{the smallest power of two greater than }c+8+|x| \]
strings of the appropriate form.  This number grows linearly (if sporadically) with $|x|$, while the number of halting programs of that form, the $|x|$th Catalan number, grows exponentially with $|x|$: a contradiction.
\QED
\fi

\medskip

\begin{cor} For every universal self-delimiting Turing machine $U$,  $1>\Omega_U> \zeta_U> \Omega_U/2>0.$
\end{cor}
\prf We have:
$2^{-\floor{\log_2(n)}} \geq 2^{-\log_2(n)} > 2^{-\floor{\log_2(n)}-1},$
where equality holds only when $n$ is a power of two, so the strict inequalities hold true because of  Proposition~\ref{zetaprefixmachine} and Fact~\ref{integerpower}.
\QED

\medskip

\begin{thm}
\label{naturalhalt}
The zeta number  $\zeta_U$ of a universal self-delimiting Turing machine $U$ is random.\end{thm}
\prf We define the machine $C$ as follows: on a string $w$, $C$ will try to compute $U(w)=y$, then continue by enumerating enough elements $\bin(n_{1}), \bin(n_{2}), \ldots , \bin(n_{k}) \in \dom (U)$ such that 
$\sum_{i=1}^{k} 1/n_{i} > 0.y$ and output $C(w) = \bin (j)$, where $j$ is the minimum positive integer
not in the set $\{n_{i} \mid 1\leq i \leq k\}$. If the computation $U(w)$ doesn't halt or the enumeration
fails to satisfy the above inequality, then $C(w)$ is undefined.

First we note that $C$ is a self-delimiting Turing machine  as $\dom (C) \subset \dom (U)$.
Secondly, if $C(w)$ is defined and $U(w')=U(w)$ with $|w'| = H_{U} (U(w))$, then $C(w) = C(w')$, hence
\begin{equation}
\label{CU}
H_{C}(C(w)) \leq |w'| \leq H_{U}(U(w)).
\end{equation}
\noindent Thirdly, because $U$ is universal, $H_{U}(x) \leq H_{C}(x) + {\rm const}_{C}$, for some 
${\rm const}_{C}$ and all strings $x$.

Finally, 

\begin{equation}
\label{approxm}
\zeta_{U} \leq 0.\zeta_U [m+1] + 2^{-m-1}.
\end{equation}
\noindent 
Given  $U(w) = \zeta_U [m+1]$ we observe that

\begin{equation}
\label{highH}
H_{U}(C(w)) > m.
\end{equation}
\noindent Indeed, if $C(w) = U(\bin (j))$, in view of (\ref{approxm}), we have:

\[
    1/j = 2^{-\log_2(j)} > 2^{-\floor{\log_2(j)}-1} \geq 2^{-m-1}.
\]

Using in order the inequality (\ref{highH}),  the universality of $U$, and (\ref{CU})  we get the folowing inequalities:\\[-3ex]
\alx{
m & <  H_{U}(C(w))\\
& \leq H_{C} (C(w)) + {\rm const}_{C}\\
& \leq H_{U}(U(w)) + {\rm const}_{C}\\
& = H_{U}( \zeta_U [m+1] ) + {\rm const}_{C},
}
\noindent proving that $\zeta_{U}$ is random.
\QED

\if01
Given $m+1$ bits of  $\zeta_U$, $\zeta_U [m+1]$, we run programs in parallel on $U$ and sum the weights of the halting programs until we can account for the given bits.  Since
\[
    1/j = 2^{-\log_2(j)} > 2^{-\floor{\log_2(j)}-1},
\]
the longest of the programs $\bin(j)$ such that $|\bin(j)| \leq m$ would affect at least the last bit, so we know that no more of these programs can halt.

We now take the outputs of all the programs that have halted and produce a string not in that set.  The complexity $H_{U}$ of that string is greater than $m$; the procedure is effective, thus, there exists a partially computable function $\Psi$ (mapping binary strings into binary strings)  with the property that
\[
    H_{U}(\Psi(\zeta_U [m+1])) > m.
\]
Consider the self-delimiting Turing machine $C(w) = \Psi (U(w))$ and note that for each string $x$,
\[ H_{C}(\Psi (x)) \le H_{U}(x).\]
By the universality of $U$ applied to $C$ we get a constant $c_{\Psi}$ such that for every string of the form $ x=\zeta_U [m+1]$ the following inequalities hold true:

\[ m < H_{U}(\Psi (x)) \le H_{C}(\Psi (x)) +c_{\Psi}\le H_{U}(x) +c_{\Psi},\]
hence
\[
    H_{U}(\zeta_U [m+1])> m-(c_\Psi+1),
\]
which shows that $\zeta_{U}$ is random.
\QED
\fi

\medskip

It is clear that $\Omega_M$ can be defined for every Turing machine, much in the same way as $\zeta_{M}$. Consequently, the zeta classification of Turing machines can be paralleled with:

\begin{defn} {\rm (Omega classification of Turing machines). According to the {\it Chaitin (Omega) halting probability}, Turing machines can be classified into the following three classes:
\begin{itemize}
\item {\it Omega divergent Turing machines}: those machines $M$ for which  $\Omega_M = \infty$,
\item {\it Omega convergent Turing machines}: those machines $M$ for which  $\Omega_M < \infty$,
\item {\it Omega  Turing machines}: those machines $M$ for which  $\Omega_M \le 1$.
\end{itemize}
}
\end{defn}

\medskip

Every self-delimiting Turing machine is an Omega Turing machine, but the converse implications is false. A natural question arises:  do the zeta and Omega classifications coincide?

\medskip

\begin{fact}
\label{hier}
{\rm a)} For every Turing machine $M$, $\zeta_{M} < \infty$ iff $\Omega _{M} < \infty$, hence the classes of  zeta divergent (convergent) Turing machines coincide. {\rm b)} If $\Omega _{M} \le 1$,  then $\zeta_{M} \le 1$, but there exists a tuatara machine $T$ such that $\Omega_T > 1$, hence the class of Omega Turing machines is strictly included in the class of tuatara machines.
\end{fact}
\prf  The equivalence a) is obvious as well as the fact that for every Turing machine $M$, $\zeta_{M} \le \Omega_{M}$. Finally, let $T$ be the Turing machine defined as follows: $T(0^{i}1)  = T(10) = 1$, for all $i\ge 0$. It is easy to see that $\Omega_{T} = 1 + 1/2 > 1> \zeta_{T}.$
\QED  

\medskip

\begin{thm}
\label{tuatarasd}
For each tuatara machine $V$ there effectively exists a  self-delimiting Turing machine $C$ such that $\Omega_{C}=\zeta_{V}$. If $V$ is tuatara universal, then $C$ can be taken to be a universal self-delimiting Turing machine.
\end{thm}
\prf   A real  $\alpha \in [0,1]$ is c.e.\
iff there  effectively exists a  self-delimiting Turing machine $C$ such that $\alpha = \Omega_{C}$ (see Theorem~7.51  in \cite{cris02}). The first part of the theorem now follows because $\zeta_{V}$ is c.e. (see Proposition~\ref{zetaprefixmachine}).

The second part of the theorem follows from Theorem~\ref{cerand=haltprob} and
Theorem~\ref{naturalhalt}.
\QED

\medskip

We can prove directly Theorem~\ref{tuatarasd}. To this aim  we need the  Kraft-Chaitin Lemma, see  \cite{cris02}:

\begin{lem}
\label{cepfset}
Given a computable enumeration of positive integers $n_i$ such that $\sum_i 2^{-n_i}\leq ~1$, we can effectively construct a prefix-free set of binary strings $\{x_i\}$ such that $|x_i|=n_i$.
\end{lem}

\medskip

We can now present a direct proof of  {\bf Theorem~\ref{tuatarasd}}:
Given a computable enumeration of positive integers $m_i$, we can write $1/m_i$ as a possibly infinite sum of reciprocals of powers of 2.  We can then lay these out on a grid and enumerate the non-zero elements along each diagonal.  For example, given the enumeration $\{1,2,3,4,5,6,\ldots\}$ the grid would be as follows:

\begin{center}\begin{tabular}{llllll}
1/2&= \phantom{,} 1/2&+ \phantom{,} 0&+ \phantom{,}0&+ \phantom{,}0&+  $\cdots$\\
1/3&=  \phantom{,} 1/4&+ \phantom{,} 1/16&+ \phantom{,}1/64&+ \phantom{,} 1/256&+ $\cdots$\\
1/4&=  \phantom{,} 1/4&+\phantom{,} 0&+\phantom{,} 0&+\phantom{,} 0&+ $\cdots$\\
1/5&= \phantom{,} 1/8&+\phantom{,} 1/16&+\phantom{,} 1/128&+\phantom{,} 1/256&+ $\cdots$\\
1/6&= \phantom{,} 1/8&+\phantom{,} 1/32&+\phantom{,} 1/128&+\phantom{,} 1/512&+ $\cdots$\\
$\ldots$
\end{tabular}\end{center}

The diagonal enumeration, taking diagonals from lower left to upper right, would be $\{1/2, 1/4, 1/4, 1/16, 1/8, 1/64, 1/8, 1/16, 1/256, \ldots\}$.  Since this enumeration is also computable, we can apply Lemma \ref{cepfset} to get a c.e.\ prefix-free set $S$.  

Let $\{m_i\}$ be an enumeration of dom($W$) and derive $S$ as above.  Define dom($V$)=S, hence $\zeta_W = \Omega_V$.

 \QED

\medskip

\if01
{\bf ??? Alternative proof for the first part:} Let $\zeta_V = 0.z_1z_2\cdots$ If all the programs $\{ V(\bin(n_1)), \ldots, V(\bin(n_{f(k)})) \}$ stop, where
\[ \sum_{i=1}^{f(k)}\frac{1}{n_i} \geq 0.z_1\ldots z_k, \]
then for $i\leq 2^k$ all programs $\bin(i)$ such that $V(\bin(i))$ halts are in the set
\[\{\bin(n_1), \ldots, \bin(n_{f(k)})\}.\]
\fi

\bigskip

\noindent  We have seen that every self-delimiting Turing machine
is a tuatara machine (Proposition~\ref{zetaprefixmachine}), but the converse is not true (Fact~\ref{hier}, b)). Another example follows.

\medskip

\begin{example}\label{prodmachine}
Given a self-delimiting Turing machine $C$ we construct a new machine $\Pi_{C}$ {\rm (}which we call a {\rm product machine}{\rm)}, such that
\pagebreak[1]
\[\dom(\Pi_{C})=\{p_1p_2\cdots p_n\mid \mbox{\rm bin}^{-1}(p_1) \leq \mbox{\rm bin}^{-1}(p_2) \leq \ldots \leq \mbox{\rm bin}^{-1}(p_n),\]\[ p_i \in \dom(C), 1\le i\le n \}, \]
and
\[\Pi_{C}(p_1p_2\cdots p_n) = C(p_{1})C(p_{2}) \cdots C(p_{n}).\]

Clearly, $\Pi_{C}$ is not  self-delimiting, but
\if01
 To prove this let us consider  a program $t$ that is known not to halt on $C$, with no halting prefixes or extensions. Construct the following self-delimiting Turing machine $C'$: $\dom(C')
= \dom (C) \cup \{t\}$, $C'(w) = C(w)$, for $w\in\dom (C)$, $C'(t)=\lambda$.  The domain of $C'$ is prefix-free since each element has exactly one copy of $t$ at the very end, and no element of $\dom(U)$ is a prefix or extension of $t$.
\fi
\[ 0< \Omega_{\Pi_{C}} = \prod_{p \in \fdom(C)} \frac{1}{1-2^{-|p|}} \leq 1.\]
\end{example}

\medskip

\noindent {\bf Comment}. The zeta number can be easily extended to Turing machines working on an arbitrary finite alphabet: we simply replace the computable bijection $\bin$ with the quasi-lexicographical enumeration of strings over the given alphabet (see more in \cite{cris02}). Because the strings in the domain of the Turing machine do not use any of the new symbols, the new bijection
maps them to a much smaller subset of the natural numbers, and every binary Turing machine becomes convergent/tuatara when thought of in the class of, say, ternary/quaternary machines.

\medskip

Next we answer in the affirmative the following question: is every Omega Number  also a zeta number?  To answer, we need two simple lemmata.

\medskip

\begin{lem}
\label{ineq} If $M, M' \ge 2$ are integers, and $q>0$ is a rational such that $1/M \le q < 1/(M-1)$
and $1/M' \le q - 1/M < 1/(M'-1)$, then $M < M'$.
\end{lem}

\medskip

\begin{lem}
\label{distinctunitlarge} Fix an integer $N\ge 2$. Then, every rational  can be effectively written as a finite sum of distinct unit fractions whose denominators are all greater than or equal to $N$.
\end{lem}
\prf Let $H_{i,j} = \frac{1}{i} + \frac{1}{i+1} + \cdots + \frac{1}{j}\raisebox{.5ex},$ for $i\le j$. Fix the rational $q$. As for every $i\ge 1$,  $\lim_{j \rightarrow \infty} H_{i,j} = \infty$, given $N \ge 2$ we can effectively find an integer $k\ge 0$ (depending on $N$) such that
\[H_{N, N+k} \le q < H_{N, N+k+1}.\]
Put $q' = q - H_{N, N+k}$ and note that
\begin{equation}
\label{1}
0\le q' < \frac{1}{N+k+1}\raisebox{.5ex}.
\end{equation}
We apply now the greedy algorithm for representing $q'$ as an Egyptian fraction (i.e. as sum of distinct unit fractions, see \cite{DE}) and we show that the denominators of all unit fractions will be larger or equal to $N$. First we get an integer $M' \ge 2$ such that
\begin{equation}
\label{2}
\frac{1}{M'} \le q' < \frac{1}{M' -1}\raisebox{.5ex},
\end{equation}
and we note that in view of (\ref{1}) and (\ref{2}) we have $M' > N+k+1$. We continue with the greedy algorithm
\begin{equation}
\label{3}
\frac{1}{M''} \le q' - \frac{1}{M'}  < \frac{1}{M'' -1}\raisebox{.5ex},
\end{equation}
and we apply Lemma~\ref{ineq} to (\ref{2}) and (\ref{3}) to deduce that
$M'' >M'$. The algorithm eventually stops because the greedy algorithm always stops over the rationals as
 the numerator decreases at each step (it must eventually reach 1, at  which point what remains is a unit fraction, and the algorithm terminates).
\QED
\medskip

\begin{thm}For each  self-delimiting Turing machine $C$ there effectively exists a  tuatara machine $V$ such that $\zeta_{V}=\Omega_{C}$.
\end{thm}
\prf We start with the expansion of $\Omega_{C} = \sum_{i\ge 1}
2^{-|x_{i}|}$, where $x_{1}, x_{2}, \ldots$ is a c.e.\ enumeration of $\dom(C)$ and we use Lemma~\ref{distinctunitlarge} to produce a c.e.\ enumeration of non-negative distinct integers $n_{1},n_{2}, \ldots$ from the representations as sum of distinct unit fractions of the terms $2^{-|x_{1}|}$, $2^{-|x_{2}|}, \ldots$, and finally we define $V(\bin(n_{i}))=\bin(n_{i})$.
\QED

\medskip

Actually, we can describe a more precise simulation of a self-delimiting Turing machine with a tuatara machine. Let $HW(p)$ be the Hamming weight of the string $p$, i.e. the number of 1 bits in $p$.

\medskip

\begin{thm}
\label{sch1}
Given a  self-delimiting Turing machine $C$ we can effectively construct a tuatara machine $V$ such that $\zeta_{V}=\Omega_{C}$. Furthermore, $\dom (V) \supset \dom (C)$, and to each string $p\in \dom (C)$ we have $HW(p)+1$ strings in $\dom (V)$, $p$ among them.
\end{thm}
\prf We define the domain of the tuatara machine $V$ to be 
\[ \dom (V) = \bigcup_{p \in \fdom(C)} X(p), \]
where $X(p)$ is the set $\{p\} \cup \{p0^i | p_i = 1\}$ and $p_i$ is the $i$th bit of $p$, numbering from the left and starting with $i=1$. We note that for each $p\in \dom(V)$ with $p_{i}=1$ we have $\bin^{-1}(p0^{i}) = 2^{i} \cdot \bin^{-1}(p)$, so for every $p\in \dom (V)$ we have:
\[\sum_{x\in X(p)} \frac{1}{\fbin^{-1}(x)} = \frac{1}{\fbin^{-1}(p)} + \sum_{i=1}^{|p|} \frac{p_i}{2^i \fbin^{-1}(p)} 
     = \frac{\fbin^{-1}(p)}{2^{|p|}\fbin^{-1}(p)} = 2^{-|p|}.\]
     
Consequently, the contribution of $2^{-|p|}$  to $\Omega_{V}$ is matched by the sum of distinct unit fractions $\sum_{x\in X(p)} \frac{1}{\fbin^{-1}(x)}\raisebox{.5ex},$  for each $p \in \dom (C)$, so $\zeta_{V}=\Omega_{C}$\footnote{For example, $X(1011)=\{1011, 10110, 1011000, 10110000\}$;
$\tfrac{1}{\fbin^{-1}(1011)}=1/27$, and $1/27+1/54+1/216+1/432 = 1/16$.}.  Furthermore, $X(p)$ has $HW(p)$ elements, and for distinct strings $p,q\in \dom(V)$, the sets $X(p)$ and $X(q)$ are disjoint, hence the unit fractions derived are mutually distinct.\QED

\medskip

\begin{sch}
\label{zetaomega}
Given a universal self-delimiting Turing machine $U$ we can effectively construct a tuatara machine $W$ universal for all self-delimiting Turing machines such that $\zeta_{W}= \Omega_{U}$.
\end{sch}
\prf In case $U=C$ is a universal self-delimiting Turing machine, the construction in the proof of Scholium~\ref{sch1} gives  a tuatara machine $W$ which is universal (but not self-delimiting) for the class of self-delimiting Turing machines. 
\QED

\medskip

\if01

\begin{sch}
\label{sch2}
Given a universal self-delimiting Turing machine $U$ with $\Omega_{U} < 1/2$, we can effectively construct a universal self-delimiting Turing machine $W$ such that $\Omega_{U}=\zeta_{W}$.
\end{sch}
\prf Because $\Omega_{U} < 1/2$, $2\Omega_{U}$ is also c.e.\ and random, so by  
Theorem~\ref{cerand=haltprob} we can construct a universal self-delimiting Turing machine $U'$ such that $\Omega_{U'} = 2\Omega_{U}$. Now define the universal self-delimiting Turing machine $U''$  by $U'' (0x) = U'(x)$ and note that $\Omega_{U''} = \Omega_{U}$ and
$\dom (U'') \subset 0\{0,1\}^{*}$. {\bf To be completed.} \QED
\fi

\medskip

Next we turn our attention to universal convergent/tuatara machines.

\medskip

\begin{thm}The sets of convergent machines and tuatara machines are c.e.\end{thm}
\prf If $M\ge 1$ is an integer and $C[M]=\{T\mid T \mbox{ is a Turing machine with } \zeta_{T}\le M\}$, then  the set of convergent machines  is $\cup_{M\ge 1} C[M]$ and $C[1]$ is the set of tuatara machines. Standard proofs (see \cite{cris02})
show that both sets are c.e. \QED

\medskip

\begin{thm}
\label{universalitytuatara}
Let $(C_i)_{i\ge 1}$ be an enumeration of tuatara machines. We define
$W(0^i1x) = C_i(x),$ for all $x \in \Sigma^*$.
Then, $W$ is a universal tuatara machine.\end{thm}
\prf First note that $0^i1\bin(n)=\bin(2^{i+1+\floor{\log_2(n)}}+n), n\ge 1$.  Now $W$ acts as follows:

\begin{equation}
\label{binextract}
 W(\bin(2^{i+1+\floor{\log_2(n)}}+n))=C_i(\bin(n)).
 \end{equation}

The machine $W$ is universal because it can
simulate any other tuatara machine with a constant prefix, and 
it is tuatara because:

\alx{
    \zeta_W &= \sum_{k \in  \Upsilon[\fdom(W)]} \frac{1}{k} \\
    &= \sum_{i\geq 1}  \sum_{n \in \Upsilon[\fdom(C_i)] } \frac{1}{2^{i+1+\floor{\log_2(n)}}+n}\\
    &= \sum_{i\geq 1}   \sum_{n \in \Upsilon[\fdom(C_i)] } \frac{1}{2^{i+1}2^{\floor{\log_2(n)}}+2^{\log_2(n)}}\\
    &\leq \sum_{i\geq 1}   \sum_{n \in \Upsilon[\fdom(C_i)] } \frac{1}{(2^i+1)2^{\log_2(n)}}\\
    &= \sum_{i\geq 1} \frac{1}{2^i+1}  \cdot  \sum_{n \in \Upsilon[\fdom(C_i)] } \frac{1}{2^{\log_2(n)}}\\
    &\leq \sum_{i\geq 0} \frac{1}{2^i+1}  \cdot \sum_{n \in \Upsilon[\fdom(C_i)] } \frac{1}{n}\\
    &\leq \sum_{i\geq 0} \frac{1}{2^i+1} \cdot \zeta_{C_i}\leq 1.\\[-2ex]
}
\QED

\medskip

\noindent {\bf Comment}. The same argument as in Theorem~\ref{universalitytuatara} shows that each $C[M]=\{T\mid T \mbox{ is a Turing machine with } \zeta_{T}\le M\}$ has a universal machine.

\medskip

\begin{thm}
\label{universalityconvergent}
There exists a universal convergent machine; furthermore, this machine can be chosen to be tuatara.
\end{thm}
\prf  If $(C^{M}_i)_{i\ge 1}$ is an enumeration of $C[M]$, then we define
$W(0^{J(i,M)}1x) = C^{M}_i(x),$ for all $x \in \Sigma^*$; here $J(i,M)= 2^{i} (2M+1)-1$. In view of (\ref{binextract}) and 

\alx{
    \zeta_W &= \sum_{k \in  \Upsilon[\fdom(W)]} \frac{1}{k} \\
    &= \sum_{i\geq 1, M\geq 1}  \sum_{n \in \Upsilon[\fdom(C^{M}_i)] } \frac{1}{2^{J(i,M)+1+\floor{\log_2(n)}}+n}\\
     &\leq \sum_{i\geq 1, M\geq 1}   \sum_{n \in \Upsilon[\fdom(C^{M}_i)] } \frac{1}{(2^{J(i,M)}+1)2^{\log_2(n)}}\\}
     \alx{
    &= \sum_{i\geq 1, M\geq 1} \frac{1}{2^{J(i,M)}+1}  \cdot  \zeta_{C^{M}_{i}}\\
    &\leq  \sum_{i\geq 1, M\geq 1} \frac{M}{2^{2^{i}(2M+1)-1}+1}\\
    &\leq  \sum_{i\geq 1, M\geq 1} \frac{1}{2^{2^{i}}+1} \raisebox{.55ex}.  \frac{M}{2^{2M}+1} <1/2.}
    
  \if01
    \alx{
    &= \sum_{i\geq 1, M\geq 1} \frac{1}{2^{J(i,M)}+1}\cdot \zeta_{C^M_i} \\
&\leq \sum_{i\geq 1, M\geq 1} \frac{1}{2^{J(i,M)}}\cdot \zeta_{C^M_i} \\
&\leq \sum_{i\geq 1, M\geq 1} \frac{M}{2^{J(i,M)}} \\
&= \sum_{i\geq 1, M\geq 1} \frac{M}{2^{2^i(2M+1)}} \\
&= \sum_{M\geq 1} M \sum_{i\geq 1} \frac{1}{\left(2^{2M+1}\right)^{2^i}} \\
&\leq \sum_{M\geq 1} M \frac{2}{2^{2M+1}} \\
&= \sum_{M\geq 1} \frac{M}{2^2M} \\
&\leq 1/2
    }
    \fi
\noindent it follows that $W$ is  tuatara and universal for the class of convergent machines.\QED

\section{Natural complexity}

Many properties can be elegantly expressed in terms of complexity. For example,  $U$ is a universal  self-delimiting  Turing machine iff  for every  self-delimiting  Turing machine $C$ there exists a fixed constant $c$, depending on $U$ and $C$, such that for every string $x\in \Sigma^{*}$,
$H_{U} (x) \le H_{C}(x) + c.$ In this spirit we present a complexity-theoretic proof of the randomness of the zeta number of a universal tuatara machine. We need first the following definition:

\medskip

\begin{defn} {\rm \cite{cs}}
{\rm  The}  natural complexity {\rm of the string $x\in\Sigma^{\ast}$ (with respect  to the tuatara machine $V$) is}
$\nabla_V(x) = \min\{n\ge 1 \mid V({\rm bin}(n))=x\}.$
\end{defn}

\medskip

\begin{fact} 
\label{utm}
{\rm  \cite{cs}} {\rm a)}  A  tuatara machine $W$ is universal  iff for every tuatara machine $V$ there exists a constant $\varepsilon$ (depending upon $W$ and $V$) such that $\nabla_{W}(x) \le \varepsilon \cdot \nabla_{V}$, for all strings $x \in \Sigma^{*}$.

{\rm b)}  A real $\alpha \in (0,1)$ is random iff there exist a universal  tuatara machine $W$ and an $ \varepsilon>0$ such that for all $n \geq 1,$ $ 2^{-n} \cdot\nabla_W(\alpha[n])\geq\varepsilon.$

\end{fact}

\medskip

\noindent {\bf Comment}. The natural complexity of a string $x$ is the position in the enumeration given by $\bin$ of the `elegant' program for $x$, denoted $x^{*} = \bin (\nabla_{W}(x))$. 
The following facts follow from the definition:

\begin{itemize}
\item for each string $x$, $W(\bin (\nabla_{W}(x)))=x$,
\item for every $j\geq 1$, if $W(\bin (j))=x$, then $\nabla_{W}(x) \leq j$,
\item for each string $x$, $x^{*}$ is the minimal (according to the quasi-lexicographical
order) input for $W$ producing $x$.
\end{itemize}
\medskip

\begin{example} For the tuatara machine constructed in the proof of Theorem~\ref{universalitytuatara}  we have: $\nabla_W(x) \leq 2^{i+1}\cdot \nabla_{C_i}(x).$
\end{example}

\medskip
\begin{thm}The zeta number $\zeta_W$ of a universal  convergent (tuatara) machine $W$ is random.\end{thm}
\prf  The proof follows the same steps as the proof of Theorem~\ref{naturalhalt}. We define the
tuatara machine $D$ acting as follows: on a string $w$, $D$ will try to compute $W(x)=y$, then continue by enumerating enough elements $\bin(n_{1}), \bin(n_{2}), \ldots , \bin(n_{k}) \in \dom (W)$ such that 
$\sum_{i=1}^{k} 1/n_{i} > 0.y$ and output $C(w) = \bin (j)$, where $j$ is the minimum positive integer
not in the set $\{n_{i} \mid 1\leq i \leq k\}$. If the computation $W(x)$ doesn't halt or the enumeration
fails to satisfy the above inequality, then $D(x)$ is undefined.

If $D(x)$ is defined, then $W(x)$ is also defined, so $D$ is tuatara. More,  $W(x) = W(x^{*})$, where $x^{*} = \bin (\nabla_{W}(x))$. It follows that $D(x) = D(x^{*})$, hence
\begin{equation}
\label{DW}
\nabla_{D}(D(x)) \leq \bin ^{-1}(x^{*}) = \nabla_{W}(W(x)).
\end{equation}

By universality of $W$ we get a constant $\varepsilon_{D}>0$ such that for all strings $x$,
\begin{equation}
\label{Wuniv}
\nabla_{W}(x) \leq \varepsilon_{D}\cdot  \nabla_{D}(x).
\end{equation}

Next we show that if $W(x) = \zeta_{W}[m]$, then 
\begin{equation}
\label{approxn}
\nabla_{W}(D(x)) > 2^{m}.
\end{equation}

\noindent Indeed, from $\nabla_{W}(D(x)) \leq 2^{m}$ it follows that if $W(\bin (j))=x$, then $1/j$
contributes towards $\zeta_{W}$, so it has to be no larger than $2^{m}$.

Using in order the inequalities (\ref{approxn}),  (\ref{Wuniv}), and (\ref{DW})  we get the folowing inequalities:\\[-3ex]
\alx{
2^{m} & <  \nabla_{W}(D(x))\\
& \leq \varepsilon_{D}\cdot  \nabla_{D} (D(x)) \\
& \leq  \varepsilon_{D}\cdot  \nabla_{W} (W(x)) \\
& = \varepsilon_{D}\cdot  \nabla_{W} (\zeta [m]), \\[-3ex]
}
\noindent proving that $\zeta_{U}$ is random.
\QED

\medskip

\if01 
First we note that in view of Theorem~\ref{universalityconvergent} we can assume that $W$ is tuatara universal.  The proof is similar to one for Theorem~\ref{naturalhalt}:  Given $m+1$ bits of  $\zeta_W$, we run programs in parallel on $W$ and sum the weights of the halting programs until we can account for the given bits.  Since
\[
    1/j = 2^{-\log_2(j)} > 2^{-\floor{\log_2(j)}-1},
\]
the longest of the programs $\bin(j)$ such that $|\bin(j)| \leq m$ would affect at least the last bit, so we know that no more of these programs can halt. We now take the outputs of all the programs that have halted and produce a string $x$ not in that set which will have the complexity  $\nabla_{W}$ greater than $2^{m}$.  Since the procedure is effective we get announced result.\QED

\begin{prop} The set of $\{\zeta_W\mid W \mbox{ is universal tuatara}\}$ is dense in $[0,1]$.
\end{prop}
\prf Indeed, if  $\zeta_W=0.z_1z_2\cdots$, $N\ge 0$, and $x$ is a binary string then $0.xz_{N+1}z_{N+2}\ldots$ is also the zeta number of a universal tuatara machine.
\QED

\medskip

\begin{cor} (Calude-J\" urgensen incompleteness)  Let $\Gamma_U(x)=2^{-|x|}\nabla_U(x)$; then $FAS$ as in G\" odel means $\exists N$ s.t. $(FAS \vdash \mbox{theorem}) \rightarrow (\Gamma_U(\mbox{theorem})\leq N).$
\end{cor}

{\bf Question:} Is $\zeta_U$ a ``real'' probability?
{\bf Answer:} Yes

$\mathbb{C}$ is the $\sigma$-algebra generated by cylinders.

\begin{thm}There exists a bijective correspondence between the probabilities defined on the $\sigma$-algebra  $\mathbb{C}$ and the functions $h: \{0,1\}^* \rightarrow [0,1]$ having the following two properties:
\begin{enumerate}
\item[1) ] $h(\lambda)=1$,
\item[2) ] $h(x) = \sum_{i=1}^{Q} h(xa_{i})$, for all $ x \in \{0,1\}^{*}$.
\end{enumerate}
\end{thm}

To get $\zeta$ we use the function $h(\bin(2^n))=h(\bin(2^n)+1)= \cdots
h(\bin(2^{n+1})-1) = 2^{-n}$. The measure of the cylinder $x\{0,1\}^*$
is $h(\bin^{-1}(x))$.
\fi

\medskip

Chaitin considered LISP program-size complexity \cite{greglisp} and found that the number of characters required in a program to produce the first $n$ bits of LISP's halting probability was asymptotic to $n/\log_2 {\mbox{\rm (number of characters)}}$.  This is the first use we know of where an author has considered the asymptotic randomness of a string and the idea that the lower bound on the complexity of prefixes of a binary sequence might be proportional to a constant less than one  times the length of the
prefix.  In this case, the constant comes from considering characters rather than bits.

Staiger \cite{St93,
St98}, Tadaki \cite{tadaki}, and Calude, Terwijn and Staiger \cite{cst} have studied the degree of randomness of sequences or reals by measuring their ``degree of compression''. Tadaki \cite{tadaki} studied the partial randomness of a generalisation of Chaitin's halting probability.  The lower bound on the complexity of successive prefixes of a random sequence is a line with slope 1.  The lower bound for the prefixes of a partially random sequence is a line with slope  $<1$.

\medskip

More precisely, following \cite{tadaki} (see also \cite{cst}), for every $ s > 0$ and universal self-delimiting Turing machine $U$ we  define the  real:\footnote{Tadaki's original notation was $\Omega_{U}^D$, where $D=1/s$.}
\[
    \Omega_{U}(s) = \sum_{p\;\in\;\fdom(U)} 2^{-s|p|}.
\]

If $0< s < 1$, then $\Omega_{U}(s) =\infty$.

\medskip

\begin{thm} {\rm \cite{tadaki}}
\label{tadaki}
For every computable $s > 1$, the  number $\Omega_{U}(s)$ is Chaitin $1/s-$random, that is, there exists a constant $c>0$ such that for all
$m\ge 1$ we have:
\[
     H_{U}(\Omega_{U}(s) [m]) \geq m/s-c.
\]
\end{thm}

\medskip

An earlier result in  algorithmic information theory states that
 there is no infinite sequence whose prefixes  have all  maximal $K_{T}$ complexity  (see more in \cite{cris02}). To state this result more precisely we fix a universal Turing machine $T$ and denote by $K_{T}$ the induced plain complexity.
 
\medskip

\begin{thm} {\rm (\cite{ch1,mar2})}
\label{gregml}
For every $c>0$ the set $\{ \alpha \in (0,1) \mid K_{T}(\alpha [m]) \geq m-c, \mbox{  for all } m\ge 1\}$ is empty.
\end{thm}

\medskip

Theorem~\ref{gregml} has given rise to alternate definitions of random sequences with respect to the plain complexity \cite{dh} and a characterisation of random reals:  {\it the real $\alpha \in (0,1)$ is random iff $K_{T}(\alpha [m]) \geq m-K_{T}({\rm bin}(m))-c$, for all $m\ge 1$}, \cite{milleryu}. We are not going to pursue this line here, but instead we will study the validity of Theorem~\ref{gregml} for partial randomness.

\medskip

Given a computable $s>1$, we will investigate 
 reals  $\alpha^s \in (0,1)$ for which there is  constant $c>0$ such that for every $m\ge 1$:
\begin{equation}
\label{partialKrand}
   K_{T}(\alpha^{s}[m]) \geq m/s-c.
\end{equation}

\noindent A real $\alpha^{s}$ satisfying the inequality (\ref{partialKrand}) will be called ``$1/s-K$-random''.

\medskip

 Random reals satisfy (\ref{partialKrand}).  We will investigate  some other examples 
 of  ``$1/s-K$-random'' reals.

\begin{example} 
\label{xx}
Let $T$ be a universal Turing machine and define $M(xx)=T(x)$, for every string $x$. The  zeta number of $M$ is $1/2-K$-random.
\end{example}

\medskip
One particularly simple self-delimiting Turing machine is  Barker's language Iota \cite{BI}.  The simplest way to define Iota is in terms of Church's $\lambda-$calculus: the universal basis $\{S=\lambda xyz.xz(yz), K=\lambda xy.x\}$ suffices to produce every lambda term, but for universality it is not necessary to have two combinators.  There are one-combinator bases, known as {\em universal combinators}.   Iota is a very simple universal combinator, $\lambda f.fSK$, denoted $0$.  To make Iota unambiguous, there is a prefix operator, $1$, for application.

The construction is essentially a very stripped-down version of LISP with only one atom, 0; since the atom takes a single input, we can represent the open parenthesis with 1, and we note that closing parentheses are unnecessary.

\if01
Valid Iota programs are pre-order traversals of full binary trees.  The number of full binary trees with $n$ leaves is $C_{n-1}$, the $(n-1)$st Catalan number ($C_n \sim \frac{4^n}{\sqrt{\pi n^3}}$), and any traversal of such a tree with $n$ leaves will be $2n-1$ bits long.  Then, the sum over all trees using the natural measure is 1:

\begin{equation}
\label{iota=1}
\sum_{n>0} \frac{C_{n-1}}{2^{2n-1}} = 1. 
\end{equation}

That is, every infinite binary sequence (except for a set with measure zero) begins with the pre-order traversal of some full binary tree.
\fi

\medskip

\begin{example} 
\label{iotakrand}
The zeta number of Iota is at least  Chaitin $1/193-$random. 
\end{example}
\prf
The traditional representation of $F$ and $T$ in combinatorial logic is
$K=F$ and $KI=T$.  In Iota, these are represented by the strings $1010100$
and $10100$, respectively.  We can encode bit strings as lists $\langle head,
tail\rangle$, where the pairing operator  $\langle-,-\rangle$  is the lambda-calculus term
$P=\lambda xyz.zxy$. In Iota, this operator is encoded by the 184-bit
string

\begin{quote}
P=1110101010011101010100110101001010101001110101010011010100101\\
\phantom{P=}0100111010101001101010010101010011101010100110101001101010100\\
\phantom{P=}1001110101010011010100101010010011101010100110101001010100100
\end{quote}

\noindent  It has the property that $1F11Pxy = x$, while  $1T11Pxy = y$, so we can
extract the head and the tail of the list.  We can distinguish a list from a Boolean value, so by terminating the list with $F$, we can know when we have read the whole string.  Each bit in the list requires two applications (one to apply $P$ to the head, and another to apply the result to the tail), the pairing operator itself, and a Boolean value.  The longest this can possibly be is 2+184+7 = 193 bits.  We can write a program that will read a bitstring $x=F^iTp$, where $p$ is any string of bits, and return $C_i(p)$, where $C_i$ is the $i$th self-delimiting Turing machine in an enumeration of the set.  
The zeta number will be at least Chaitin $1/193-$random, because it takes no more than $193n+c$ bits to ouptut $n$ bits of $\Omega_U$ for any universal self-delimiting Turing machine $U$. 
Therefore  the zeta number of Iota itself is at least Chaitin $1/193-$random. 
\QED

\bigskip

\noindent {\bf Comment}. A sharper result will be presented in Example~\ref{iotakrandc}.
There are much better encodings available in Iota than the naive one above. We conjecture that, in fact, Iota's zeta number is ``more random'' (see our list of open questions at the end of this paper).

\if01
Since the Catalan numbers are asymptotic to $4^n/\sqrt{\pi n^3}$, the terms in the sum    $1-\sum_{i=1}^n C_{i-1}/2^{2i+1}$ are asymptotic to $1/2\sqrt{\pi n^3}$; hence, if we know the halting status of all programs of length up to about $3n$ bits, we can calculate $n$ bits of $ \zeta_{U}$.
\fi

\medskip

We continue with a more general construction. Tadaki's generalization of Chaitin's halting probability (see \cite{tadaki}) is a zeta function.  Zeta functions appear as partition functions and in expectation values in statistical systems, and the parameter $s$ corresponds to an inverse temperature.  The partition function for a statistical system $X$ has the form
\[ Z(s) = \sum_{x \in X} e^{-sH(x)}, \]
where $H$ is the energy (Hamiltonian) of the state $x$, and $s$ is inversely proportional to the temperature of the system.\footnote{The notation $H$---used only in this  motivational part---does not denote the program-size complexity, although it is not too far away from it.}  An observable is a function $\kappa:X
\rightarrow \RR$.  The average value of the observable for a system at equilibrium is
\[ \langle \kappa \rangle (s) = \frac{\sum_{x \in X} \kappa(x) e^{-sH(x)}}{Z(s)} \raisebox{.5ex}. \]
The partition function acts like a normalization constant.

Taking $X$ to be the set of programs, we let the ``energy'' of a program be its length.  The partition function becomes 
\[ Z(s) = \sum_{p \in X} e^{-s|p|}.\]
We can recover the base 2 if we let $s=s' \ln 2$:
\[ Z(s' \ln 2) = \sum_{p \in X} 2^{-s'|p|}.\]
Taking $X$ to be prefix-free guarantees that the partition function converges at $s'=1$ by the Kraft-Chaitin Lemma; however, the function converges for any subset of $\Sigma^*$ when $s'>1$.  When $X=\Sigma^*$, the set of all binary strings, $Z(s' \ln 2) = 1/(1-2^{-s'+1}).$ 

\medskip

We now define our observable to be the halting function of the Turing machine $T$: $\kappa_T(p) = 1$ if $T$ halts on $p$, 0 otherwise.  The probability that a program will halt is then
\[ \langle \kappa_T \rangle (s) = \frac{\sum_{p \in X} \kappa_T(p) 2^{-s'|p|}}{Z(s)} = \frac{\Omega_T(s')}{Z(s)} \raisebox{.5ex}. \]

All of this passes over nicely to the zeta number.  We let $X=\NN$ and let the ``energy'' of $n \in \NN$ be $\ln n$.  The partition function becomes
\[ Z(s) = \sum_{n \in \NN} e^{-s \ln n} = \sum_{n \in \NN} n^{-s} = \zeta(s), \]
the Riemann zeta function.  
\medskip

We define the ``zeta function of $T$'' to be 
\[ \zeta_T(s) = \sum_{n \in \NN} \kappa_T(\bin(n)) \cdot n^{-s} = \sum_{n\in  \Upsilon[\fdom(T)]} n^{-s}, \]
and the probability that a program will halt on $T$ is
\[ \langle \kappa_T \rangle (s) = \frac{\zeta_T(s)}{\zeta(s)}  \raisebox{.5ex}.  \]

\medskip

Given a Turing machine $M$ (which may or may not be self-delimiting), we define 
``the halting probability of $M$ at $s$'' to be

\[
    \langle \kappa_M \rangle (s) = \left(\sum_{p\;\in\;\fdom(M)} 2^{-s|p|}\right)\mbox{\huge /}\left(\sum_{q\;\in\;\NN} 2^{-s|\fbin(q)|}\right)
 =  (1-2^{-s+1}) \sum_{p\;\in\;\fdom(M)} 2^{-s|p|}.\]

\medskip

\begin{fact}    For real $s>1$ and universal $T$, $0 < \langle \kappa_T \rangle (s) < 1$.\end{fact}
\prf
Since $T$ is a universal Turing machine, then there must be some integer $q$ such that $\bin(q)\not\in\dom(T)$.  Therefore the numerator, which sums only over those $q$ such that $\bin(q) \in \dom(T)$, is smaller than the denominator, which sums over all positive natural $q$.  Since there must be at least one program that halts, the numerator is positive.

\QED

\bigskip

\begin{thm}
\label{parrand}
For every  computable real  $s>1$ and universal $T$, $\langle \kappa_T \rangle (s)$ is $1/s-K-$random.\end{thm}
\prf
Given the first $m+\lfloor \log_2(1-2^{-s+1})\rfloor $ bits of $\langle \kappa_T \rangle (s)$, we can compute the halting status of all programs $p\in\dom(T)$ such that $|p|<m/s$.  Then, there is a computable function $\Psi$ that, given $\langle \kappa_T \rangle (s)  [m+\floor{\log_2(1-2^{-s+1})}] $, produces a string not in the output of those programs, hence

\[
    K_{T}(\langle \kappa_T \rangle (s) [m]) \geq m/s-(c_\Psi+\lfloor \log_2(1-2^{-s+1})\rfloor).
\]
\QED

\medskip

\noindent {\bf Comment}. 
a) The number $\langle \kappa_T \rangle (s)$ is  a halting probability (see \cite{cris02}).  One particularly nice value is $s=2$ where $\sum_{n> 0}2^{-2\floor{\log_2(n)}} = 2$.  With reference to Example~\ref{xx}, 
if $T=U$ is  self-delimiting, then $\Omega_M = \Omega_M(1)$ is Chaitin $1/2-$random:

\[\Omega_M(1)  = \sum_{x \in \fdom(M)} 2^{-|x|}
    = \sum_{x \in \fdom(U)} 2^{-2|x|}
    = \Omega_U(2)
    = 2 \langle \kappa_U \rangle(2).\]

b) Theorem~\ref{parrand}  shows  a property true for partial random reals, but not for random reals, cf.  \cite{cris02}.  An opposite phenomenon was described in  \cite{cst}.  The following characterisation of random reals is no longer true for partial random reals:
{\it A real  $\alpha \in (0,1)$ is random iff there exist a constant $c \ge 0$ and an infinite computable set $M\subseteq \NN$ such that $H_{U}(\alpha[n]) \ge n-c$, for each $n\in M$.}

\bigskip

Obviously, if $\alpha$ is $1/s-K$-random, then it is also Chaitin $1/s-$random. 

\medskip

\begin{cor}If $U$ is a universal self-delimiting Turing machine, then for every computable real $s>1$,  $\langle \kappa_U \rangle (s)$ is Chaitin $1/s-$random.\end{cor}

\medskip

Note that in this case, $\langle \kappa_U \rangle (s)$ is just a computable factor times $\Omega_{U}(s)$.

\medskip

Furthermore, the converse implication is false:

\begin{prop}
\label{hnot=k}
There exists a Chaitin $1/2-$random real which is not $1/2-K$-random.\end{prop}
\prf Let $K=K_{T}$, where $T$ is a universal Turing machine, and let
$\alpha = 0.x_{1}x_{2}\cdots x_{n} \cdots$ be Chaitin $1-$random. On one hand, the real $\alpha$ is not $1-K$-random, cf. \cite{cris02}. On the other hand, the number $\beta=0.0x_{1}0x_{2}\cdots 0x_{n} \cdots$ is Chaitin $1/2-$random; if $\beta$ were
$1/2-K$-random, then $\alpha$ would be $1-K$-random, a contradiction. Indeed, for all $n\ge 1, K_{T}(0x_{1}0x_{2}\cdots 0x_{n}) \le K_{F \circ T}(0x_{1}0x_{2}\cdots 0x_{n}) + c' \le K_{T}(x_{1}x_{2}\cdots x_{n}) + c'$, where $F\circ T (y) = 0x_{1}0x_{2}\cdots 0x_{n}$ whenever $T(y) = x_{1}x_{2}\cdots x_{n}$.

\QED

\medskip

\begin{lem}
\label{elimconstant}
Let $\alpha \in (0,1)$. If there exist two integers $c, N\ge 0$ and a real $a\in (0,1]$  such that for all $m >N$ we have $K_{T}(\alpha [m]) \ge a\cdot m-c$, then we can find a constant $b\ge 0$ such that $K_{T}(\alpha [m]) \ge a\cdot m -b$, for all $m\ge 1$.
\end{lem}
\prf Put $b = \max_{1\le i\le N} \max \{0, a\cdot i - K_{T}(\alpha [i])\}+c$.

\QED

\medskip

\if01
We now define a new form  of partial randomness by requiring that the real is as close as we wish to being random, without necessarily being random. Here is the formal, more general, definition:

\medskip

\begin{defn} {\rm  Let $s>1$ be computable. We say that a real number $\alpha \in (0,1)$ is } asymptotically $1/s-K$-random {\rm (}asymptotically Chaitin $1/s-$random{\rm )} {\rm if
for every computable real $t > s > 1$ there exists a constant $c_{t}\ge 0$ such that for all $m\ge 1$ we have $K_{T}(\alpha [m]) \ge m/t-c_{t}$ ($H_{U}(\alpha [m]) \ge m/t-c_{t}$).}

{\rm If $s=1$, then  $\alpha$ is called} asymptotically $K$-random {\rm (}asymptotically Chaitin random{\rm )}.
\end{defn}

\medskip

Again, every asymptotically $1/s-K$-random ($K$-random) real is asymptotically Chaitin $1/s-$random (Chaitin random). 
In contrast with Proposition~\ref{hnot=k}, following \cite{St06} we have:

\medskip

\begin{thm}
\label{h=k}
Let $s\ge1$ be computable. Then, a real $\alpha$ is asymptotically $1/s-K$-random iff
$\alpha$ is asymptotically Chaitin $1/s-$random.
\end{thm}
\prf Following \cite{St02} we define $\underline{k}(\alpha) =\liminf_{n\rightarrow \infty} \frac{1}{n} K(\alpha_{1}\ldots \alpha_{n})$. Then, for every  $s\ge1$ computable   we have: a real $\alpha$ is asymptotically $1/s-K$-random iff  $\underline{k}(\alpha)\ge 1/s$ iff
$\alpha$ is asymptotically Chaitin $1/s-$random.

\QED

\medskip
\fi

We now define a new form  of partial randomness by requiring that the real is as close as we wish to being (partially) random, without necessarily being random. 
Following \cite{Ryabko,St02} we define the lower asymptotic complexity  $$\underline{k}(\alpha) =\liminf_{n\rightarrow \infty} \frac{1}{n} K(\alpha_{1}\ldots \alpha_{n}).$$

\medskip

Following \cite{St06} we have:

\medskip

\begin{thm}
\label{h=k}
Let $s\ge1$ be computable. Then, for a real $\alpha$, the following statements are equivalent:

\begin{enumerate}
\item[{\rm 1)}]  We have: $\underline{k}(\alpha)\ge 1/s$.
\item[{\rm 2)}]  For every computable real $t > s > 1$, $\alpha$ is  $1/t-K$-random.
\item[{\rm 3)}]  For every computable real $t > s > 1$,  $\alpha$ is  Chaitin $1/t-$random.
\end{enumerate}
\end{thm}
 \prf Conditions 2) and 3)  are equivalent because of the asymptotics. The equivalence with 1) can be verified by elementary calculus.

\QED

\medskip

\begin{defn} {\rm  Let $s>1$ be computable. We say that a real number $\alpha \in (0,1)$ is } asymptotically $1/s$-random {\rm  if one of the equivalent conditions in Theorem~\ref{h=k} is satisfied.}
{\rm If $s=1$, then  $\alpha$ is called} asymptotically random.
\end{defn}

\medskip

The notion of asymptotic $1/s$-randomness   induces a strict hierarchy on $s>1$. 
We need the following result (for the definition of the Hausdorff dimension see  Falconer~\cite{Falconer}):
 
 \medskip
 
 \begin{thm} {\rm \cite{Ryabko}}
 \label{haus}
 Let $\dim_{H}$ be the Hausdorff dimension and let $s>1$ be computable.  Then:
 \[ \dim_{H}(\{\alpha \in [0,1] \mid \underline{k}(\alpha) \le 1/s\}) = 
 \dim_{H}(\{\alpha \in [0,1] \mid \underline{k}(\alpha) = 1/s\}) = 1/s.\] 
\end{thm}

\medskip

In view of Theorem~\ref{haus} we will refer only to asymptotic  $1/s$-randomness (without mentioning $K$ or $H$).  Consequently, using Theorems~\ref{h=k} and \ref{haus} we get:

\begin{cor} The notion of asymptotic  $1/s$-randomness real induces a strict hierarchy for $s>1$.
\end{cor}

\medskip

 \begin{example} 
\label{iotakrandc}
The zeta number of Iota is at least  $1/194-K$-random and at least asymptotically $1/193$-random.
\end{example}
\prf
Since we know where the encoded bit string ends, Iota can
simulate an arbitrary universal Turing machine, not just a
self-delimiting one.  For any $s>1$   we can print $m$ bits of $\zeta_U(s)$
with at most $193m+c$ bits.  So the zeta number of Iota is at least $1/194-K$-random and at
least asymptotically $1/193$-random.

 \QED

\medskip

Given an arbitrary Turing machine $M$, we define ``the {\em natural} halting probability at $s$'' to be

\[
\langle \kappa^n_M \rangle (s) = \left( \sum_{q  \in \Upsilon[\fdom(M)]} q^{-s} \right)
                \mbox{\huge /}
                \left(\sum_{q\;\in\;\NN} q^{-s}\right)= \zeta_{M}(s) / \zeta(s),
\]

\medskip

\noindent where we have added a superscript to $\kappa$ to distinguish it from the Tadaki-Chaitin case.

Next, we can define the set
\[
    P=\{p_i \mid \bin(i) \in \dom(M)\},
\]
where $p_i$ is the $i$th prime in increasing order, and  the set

\[
    S= \{n \mid  \mbox{ all prime factors of }n   \mbox{  are in  } P\}.
\]

\medskip

The set $\bin(S)$ is the domain of a Turing machine $R(M)$ ({\em prime product machine})  that performs the following steps on an input $x\in \Sigma^{*}$:
\begin{enumerate}
\item Compute $n = \bin^{-1}(x)$.
\item Compute the prime factors $p_i$ of $n$.
\item For each $p_i$, simulate $M(\bin(i))$.
\item Output the empty string.
\end{enumerate}

\medskip

Then, 
\[
    \zeta_{R(M)}(s) = \sum_{n \in S} n^{-s} = \prod_{p \in P} 1/(1-p^{-s}).
\]

\smallskip

The definition of the 
Omega number works nicely for string concatenation (see the product machine, Example~\ref{prodmachine}); the zeta number for this
machine is complicated. In contrast, 
the zeta number 
works well for integer multiplication as in the case of the prime product machine;  the Omega number of this machine is complicated.

\medskip

\begin{thm}
\label{arand} For every universal Turing machine $T$ and computable $s>1$, $\langle \kappa^n_{R(T)} \rangle (s)$  is  asymptotically $1/s$-random.
\end{thm}
\prf
The Prime Number Theorem implies that for $i>5$, $i \log(i) < p_i$. Fix a computable real $s>1$.  Given $\lfloor ms \rfloor + 1$ bits of $\langle \kappa^n_{R(T)} \rangle (s)$, we can compute the halting status of all programs $\bin(i)$ such that $p_i<2^m$. Consequently,
\alx{
    i \log(i) &< 2^m,\\
    i &< 2^m/W(2^m),\\
    |\bin(i)| = \floor{\log_2(i)} &< m-\log_2(W(2^m)) = W(2^m)/\ln (2).
}

Here, $W$ is the Lambert $W$-function, 
the inverse function of $f(x) = x e^{x}$, \cite{eww}; it has the series expansion  $W(x) = \sum_{n=1}^{\infty} \frac{(-n)^{n-1}}{n!}\, x^{n}$. Therefore, given  $\lfloor ms \rfloor + 1$ bits, we can compute the halting status of all programs whose lengths are each less than $W(2^m)/\ln (2)$.  Since
\begin{equation}
\label{lim}
    \lim_{m\rightarrow \oo} \frac{W(2^m)}{m\ln (2)} = 1,
\end{equation}
the result follows. Indeed, in view of (\ref{lim}), for each $i>5$ and $\varepsilon >0$ there exists a bound $N_{\varepsilon}$ such that $ K_{T}(\langle \kappa^n_{R(T)} \rangle (s)[m]) \geq (1-\varepsilon)m$, for every $m\ge N_{\varepsilon}$, hence in view of Lemma~\ref{elimconstant},
we can find a constant $c_{s}\ge 0$ such that for all $m\ge 1, K_{T}( \langle \kappa^n_{R(T)} \rangle (s) [m]) \geq m/s - c_{s}.$

 \QED

\medskip

\begin{cor}

\label{sdarand}
If $U$ is a universal self-delimiting machine and  $s>1$ is computable, then $\langle \kappa^n_{R(U)} \rangle (s)$ is asymptotically  $1/s-$random.
\end{cor}

\medskip

\begin{lem}
\label{logrand}
Let $\alpha \in (0,1).$ If there exist three integers $c, a, N\ge 0$ such that for all $m \ge N$ we have $K_{T}(\alpha [m+ c\lfloor \log_{2}m \rfloor ]) \ge  m - a$, then for every computable $s>1$ we can find a constant $b\ge 0$ such that 
 $K_{T}( \alpha [m]) \ge m/s -b$, for all $m\ge 1$.
\end{lem}
\prf For each computable $s>1$ we can find a constant $d\ge 0$ such that for all $m\ge 1$, $K_{T}(\alpha [m+ c\lfloor \log_{2}m \rfloor ]) \ge \frac{1}{s} (m + c\lfloor \log_{2}m \rfloor) - d$, so the required inequality follows from Lemma~\ref{elimconstant}.

\QED

\begin{thm}
\label{omegaAKrand}
 If $U$ is a universal self-delimiting Turing machine, then $\Omega_{U}$
is asymptotically $K$-random.
\end{thm}
\prf Since $\Omega_{U}$ is random, there exists a constant $c\ge 0$ such that for all $m \ge1$
$H_{U}(\Omega_{U} [m]) \ge m-c.$ 
On the other hand, there exists $a\ge0$ such that for all $m \ge 1$ we have
 $K_{T}(\Omega_{U} [m+a \lfloor \log_{2}m \rfloor ]) \ge H_{U} ( \Omega_{U} [m]) \ge m-c$, hence in view of Lemma~\ref{logrand}, for every computable $s>1$ there exists an integer  $b\ge 0$ such that for all $m\ge 1$ we have:
 $K_{T}( \Omega_{U} [m]) \ge m/s-b$. This shows that $\Omega_{U}$
is asymptotically random.

\QED

\medskip

\begin{cor}
If $U$ is a universal self-delimiting machine, then $\zeta_{U}$ is asymptotically random.
\end{cor}
\prf Use Theorem~\ref{omegaAKrand} and Scholium~\ref{zetaomega}.
\QED

\medskip

The converse implication fails to be true:

\medskip

\begin{thm}
\label{arnotrand} There is a self-delimiting Turing machine $V$ such that
$\zeta_{V}$ is asymptotically random, but not random.
\end{thm}
\prf 
Let $\overline{p}$ be a self-delimiting version of the string $p$ such that $|\overline{p}| \approx |p| + 2\log_{2}|p|$ (see for example \cite{cris02}).  Let $(C_{i})$ be a c.e.\ enumeration of all self-delimiting Turing machines and define $V(0^{i}1 \overline{p}) = C_{i}(p)$.  Clearly, there is a constant $c\ge 0$ such that  for  all $m\ge 1$, $K_{T}( \zeta_{V} [m+ 2\lfloor \log_{2}m \rfloor ]) \ge m - c$, so in view of Lemma~\ref{logrand}, $\zeta_{V}$ is asymptotically random. However, $V$ is not universal, so $\zeta_{V}$ is not random.

\QED

\medskip

\noindent {\bf Comment}. A different proof for Theorem~\ref{arnotrand} can be obtained using a non-sparse dilution, cf. Example 3.18 in \cite{St93} or Theorem 4.3
in \cite{Lutz}.

\medskip

\begin{cor} 
\label{Chatinrandnotrand}
There is a self-delimiting Turing machine $V$ such that $\zeta_{V}$ is asymptotically  random, but not random.
\end{cor}

\medskip

\noindent {\bf Comment}.   If  $ x_{1}x_{2}\cdots$ is a random sequence, then the sequence $$ x_{1}0^{\lfloor \log_{2}1\rfloor}
x_{2}0^{\lfloor \log_{2}2\rfloor- \lfloor \log_{2}1\rfloor}\cdots  
x_{n}0^{\lfloor \log_{2} n\rfloor - \sum_{i=1}^{n-1} \lfloor \log_{2} i \rfloor
} \cdots
$$  is not random, but asymptotically  random.

\medskip

\begin{lem}
\label{logbound}
For every pair of computable reals $r,t>1$ and integer $c\ge 1$ there exists a computable real $s>1$ such that for every $m\ge 1$ we have:  $(\frac{1}{s} -\frac{1}{r})\cdot m \ge \frac{c}{t} \cdot \log_{2}m$.
\end{lem}

\prf Take $\frac{1}{s} = \frac{c}{t} + \frac{1}{r}$.

\QED

\medskip

Theorem~\ref{gregml} proves that there is no infinite sequence whose prefixes  have all  maximal $K_{T}$ complexity. A similar result can be proved for program-size complexity $H_{U}$. However, this result will be  false
for asymptotic randomness.

\medskip

\begin{thm} \label{nongregml} There exists a real $\alpha \in (0,1)$ such that for every pair of
computable reals $r,t>1$ and integer $c\ge 1$, there exists an integer $b\ge 1$ such that for every $m\ge 1$,
\[H_{U} ( \alpha [m]) \ge \frac{1}{r}\cdot m + \frac{c}{t}\cdot \log_{2}m -b.\]
So, $H_{U} ( \alpha [m])$ is as close as we want, but never equal, to $\max_{|x| = m} H_{U}(x)-\mbox{O}(1)$.
\end{thm}
\prf Take an asymptotically  random real $\alpha$, consequently, for every computable $s>1$ there is a constant $a \ge 0$ such that $H_{U} (\alpha [m]) \ge \frac{1}{s} \cdot m -a,$ for all $m\ge 1$, and then use Lemma~\ref{logbound}.

\QED

\medskip

\section{Open problems}
Many interesting questions remain unsolved. For example, 
can  the machine $V$ in  Scholium~\ref{zetaomega}   be taken to be universal self-delimiting or universal tuatara?

The zeta number of Iota is at least  $1/194-K$-random and at least asymptotically $1/193$-random (Example~\ref{iotakrandc}); we conjecture that natural halting probability of Iota is asymptotically $K$-random, but not random.
 
 Let $U^{K}$ is a universal self-delimiting machine with an oracle to the Halting Problem, and $\Omega^{K}= \Omega_{U^{K}}$; $\Omega^{K} (2)$ is Chaitin $1/2-2-$random. 
Is $\Omega^{K} (2)$  random or asymptotically $K$-random?

\section*{Acknowledgement} We thank Greg Chaitin, Andr\' e Nies, and the anonymous referees for many useful comments 
and references, Nick Hay for interesting questions,  Rich Schr\"oppel for the idea on which the proof of Lemma~\ref{distinctunitlarge} is based, and Ludwig Staiger for suggesting Theorems~\ref{density} and \ref{h=k}.

\end{document}